\documentclass[prd,twocolumn,showpacs,floatfix,english,letterpaper]{revtex4}

\usepackage{amssymb}
\usepackage[dvips]{graphicx}

\usepackage[T1]{fontenc}
\usepackage[latin1]{inputenc}

\makeatletter

\providecommand{\tabularnewline}{\\}

\usepackage{babel}
\makeatother
\begin{document}

\title{More about excited bottomonium radiative decays}

\author{Randy Lewis}
\affiliation{Department of Physics and Astronomy, York University,
Toronto, Ontario, Canada M3J 1P3}

\author{R. M. Woloshyn}
\affiliation{TRIUMF, 4004 Wesbrook Mall, Vancouver,
British Columbia, Canada V6T 2A3}

\begin{abstract}
Radiative decays of bottomonium are revisited, focusing on  
contributions from higher-order relativistic effects.  
The leading relativistic correction to the magnetic spin-flip 
operator at the photon vertex is found to be particularly
important. The combination of $\mathcal{O}(v^{4})$ effects in the
nonrelativistic QCD action and in the transition operator moves previous lattice 
results for excited $\Upsilon$ decays into agreement with experiment.
\end{abstract}


\maketitle

In a recent paper \cite{lewis11} we studied bottomonium
radiative decays using lattice nonrelativistic QCD (NRQCD) \cite{latnrqcd}. It was 
shown that robust signals for transitions among the ground state and the first 
two excited states could be obtained using multiexponential fitting 
techniques \cite{constrainedfit}. 
The qualitative features of the decay amplitudes were
in agreement with phenomenological expectations but quantitatively 
the values obtained were considerably larger than those determined 
by experiment.

In Ref.~\cite{lewis11} the $b$ quarks were described using an 
$\mathcal{O}(v^{4})$ lattice NRQCD action and only the 
simplest magnetic spin-flip transition operator was used.
In this work higher-order relativistic effects in the action
and additional terms in the transition operator are investigated.
It is expected that the hindered transitions, which involve states
with different principal quantum numbers, are very sensitive to 
relativistic corrections. In comparison to Ref.~\cite{lewis11}, the additional 
effects considered in this work substantially reduce the 
amplitudes for excited $\Upsilon$ decays and bring them into 
line with experimental values.
 
The general setup for this study is the same as in Ref.~\cite{lewis11}.
The gauge fields come from a 2+1-flavor dynamical simulation done 
on a $32^3\times64$ lattice by the PACS-CS collaboration \cite{pacscs09}. 
An ensemble of
192 configurations is used. The light quark parameters are
such that the pion mass is near physical at 156 MeV. Landau link
tadpole improvement is implemented with a value of 0.8463 for the
link. The $b$-quark bare mass is 1.945 (in lattice units) and a stability 
parameter n of 4, in line with Ref.~\cite{davies10}, is used.

The three-point functions for vector to pseudoscalar transitions
are constructed using a sequential source method. 
Starting with a vector (pseudoscalar) operator at the 
source $t_{s}$, the quark propagator is evolved to a  
time $T$ at which a pseudoscalar (vector) operator is applied. This
quantity is then evolved backward in time. At intermediate times $t_{s}<t'<T$
a transition operator is inserted and evolution is continued to complete
the quark antiquark loop at the source. 

With a vector operator at the source and pseudoscalar at the sink
the three-point function is expected to have the form
\begin{eqnarray}\label{eq_threepnt}
\nonumber
G_{oo'}^{(VP)}(t';T) &=& \sum_{n,n'}c_{o}^{(V)}(n)A_{nn'}^{(VP)}c_{o'}^{(P)}(n') \\
                     && \times e^{-E_{n}^{(V)}(t'-t_{s})}e^{-E_{n'}^{(P)}(T-t')}
\end{eqnarray}
where the subscripts $o,o'$ indicate the type of operator (local or smeared)
that is used. As in Ref.~\cite{lewis11}, three types of operators are used:
a local operator, a wavefunction smeared operator \cite{davies94}
and an operator with the smearing applied twice \cite{davies10}. 
See Ref.~\cite{lewis11} for a description of the smearing function and parameters.

The overlap coefficients $c$ and simulation energies $E$ are the same ones
that appear in the two-point function. The quantity
$A_{nn'}^{(VP)}$ is the matrix element of the transition operator between
the vector state $n$ and the pseudoscalar state $n'$.  The three-point 
function with pseudoscalar source and vector sink has the same form with $V$
and $P$ labels reversed. The matrix elements $A_{nn'}^{(PV)}$ are related
to those appearing in (\ref{eq_threepnt}) by $A_{nn'}^{(PV)}=A_{n'n}^{(VP)}.$
The matrix elements can be determined by fitting the $t'$ dependence of the 
three-point function for a fixed $T$ using overlap coefficients and energies
obtained from a fit to two-point correlators.

The calculations in Ref.~\cite{lewis11} were done using an NRQCD action including
terms to $\mathcal{O}(v^{4})$. In this work $\mathcal{O}(v^{6})$ terms are
also considered. The complete action can be found, for example, 
in Ref.~\cite{meinel10}. Here we display only the spin-dependent terms 
linear in chromoelectric and chromomagnetic fields
which are relevant for subsequent discussion 
(see Ref.~\cite{meinel10} for detailed explanation of notation):
\begin{eqnarray}
\label{dH4}
\delta{H}^{(4)} &=& -\frac{c_4g}{2M_0}\mbox{{\boldmath$\sigma$}}\cdot\tilde{\bf B} \nonumber \\
            && -\frac{c_3g}{8M_0^2}\mbox{{\boldmath$\sigma$}}
               \cdot(\tilde{\bf \Delta}\times\tilde{\bf E}-\tilde{\bf E}\times
               \tilde{\bf \Delta}), \\
\label{dH6}
\delta{H}^{(6)} &=& -\frac{c_7g}{8M_0^3}\left\{\tilde\Delta^{(2)},
                        \mbox{{\boldmath$\sigma$}}\cdot\tilde{\bf B}\right\} 
                    \nonumber \\
                &&  -\frac{c_83g}{64M_0^4}\left\{\tilde\Delta^{(2)},
                     \mbox{{\boldmath$\sigma$}}\cdot(\tilde{\bf \Delta}\times\tilde{\bf E}-\tilde{\bf E}\times
                     \tilde{\bf \Delta})\right\},
\end{eqnarray}
where $\delta{H}^{(4)}$ gives the $\mathcal{O}(v^{4})$ chromomagnetic coupling and spin-orbit
terms and $\delta{H}^{(6)}$ their leading relativistic corrections. The fields are
tadpole-improved and in our calculations the action coefficients take the tree-level values. 
The transition operator should be constructed in a way which is consistent with the 
action. This can be achieved by replacing in (\ref{dH4}) and (\ref{dH6}) the SU(3) 
color electric and magnetic fields by external electromagnetic fields and the 
strong coupling constant by the charge. See for example Ref. \cite{bram06}.

For an M1 decay, the photon momentum, polarization vector and quark spin operator should be
mutually orthogonal. The explicit choice made for the operators used in our lattice simulation
is given in Table \ref{tab_ops2}. The normalization is such that in the infinite mass limit the 
transition matrix element goes to $1$, matching the nonrelativistic wavefunction overlap (see Eq.~(1) in Ref.~\cite{lewis11}).

\begin{table}

\caption{Transition operators used in calculating three-point functions. The
momentum $\mathbf{k}$ is chosen to have a component only in the one direction
and $\Delta_{1k}\equiv \left(\Delta_{1}e^{i\mathbf{k\cdot x}}+e^{i\mathbf{k\cdot x}}\Delta_{1}\right)$.}
\begin{centering}
\begin{tabular}{ccc}
\hline 
\hline
&magnetic&electric\tabularnewline
\hline 
\vspace{-.2cm}
&&\tabularnewline
\vspace{.2cm}
$\mathcal{O}(v^4)$&
$\sum_{\mathbf{x}}\sigma_{3}e^{i\mathbf{k\cdot x}}$
&
$\frac{1}{4M_{0}}\sum_{\mathbf{x}}i\sigma_{3}\Delta_{1k}$\tabularnewline
\vspace{.2cm}
$\mathcal{O}(v^6)$&
$\frac{1}{4M_{0}^{2}}\sum_{\mathbf{x}}\left\{ \Delta^{(2)},\sigma_{3}e^{i\mathbf{k\cdot x}}\right\} $
&
$\frac{3}{32M_{0}^{2}}\sum_{\mathbf{x}}\left\{ \Delta^{(2)},i\sigma_{3}\Delta_{1k}\right\} $\tabularnewline
\hline
\hline
\end{tabular}
\par\end{centering}
\label{tab_ops2}
\end{table}

Results for the three-point transition matrix elements 
are given in Table
\ref{tab_delT} for different source-sink time separations and
different values of recoil momentum.  The calculations
are done with the $\mathcal{O}(v^{4})$ action and the leading magnetic
operator $\sigma_{3}$. The values for $T-t_{s}$
equal to 19 and 27 are from Ref.~\cite{lewis11}. For the excited 
to ground state transitions, there is very good agreement between
results using different time separations. However, using a smaller
time separation  $T-t_{s}$ = 15 allows for a better determination of
the $2\rightarrow2$ transition and we use only this for the present 
study.

\begin{table*}

\caption{Three-point matrix elements from simultaneous fits to 12 correlation
functions with different source-sink separation.}

\begin{centering}
\begin{ruledtabular}
\begin{tabular}{ccccccc}
$T-t_{s}$&
$A_{11}^{(VP)}$&
$A_{21}^{(PV)}$&
$A_{31}^{(PV)}$&
$A_{21}^{(VP)}$&
$A_{31}^{(VP)}$&
$A_{22}^{(VP)}$\tabularnewline
\hline 
&
&
\multicolumn{3}{c}{momentum 0}
&
&
\tabularnewline
15&
0.916(2)&
-0.068(3)&
-0.050(3)&
0.071(5)&
0.070(7)&
0.863(67)\tabularnewline
19&
0.915(2)&
-0.068(2)&
-0.050(4)&
0.071(4)&
0.065(3)&
1.11(23)\tabularnewline
27&
0.916(2)&
-0.068(3)&
-0.051(6)&
0.071(4)&
0.062(4)&
1.9(1.8)\tabularnewline
&
&
\multicolumn{3}{c}{momentum 1}
&
&
\tabularnewline
15&
0.907(2)&
-0.060(6)&
-0.047(5)&
0.082(5)&
0.072(5)&
0.799(68)
\tabularnewline
19&
0.907(1)&
-0.062(6)&
-0.047(7)&
0.079(5)&
0.067(5)&
0.95(21)\tabularnewline
27&
0.907(2)&
-0.061(5)&
-0.048(8)&
0.079(5)&
0.066(6)&
1.6(1.5)\tabularnewline
&
&
\multicolumn{3}{c}{momentum 2}
&
&
\tabularnewline
15&
0.877(1)&
-0.031(4)&
-0.040(5)&
0.103(6)&
0.083(6)&
0.791(60)
\tabularnewline
19&
0.877(1)&
-0.031(4)&
-0.041(6)&
0.102(5)&
0.078(6)&
1.02(20)\tabularnewline
27&
0.878(2)&
-0.029(5)&
-0.039(7)&
0.100(5)&
0.068(6)&
1.0(1.6)\tabularnewline
\end{tabular}
\end{ruledtabular}
\par\end{centering}
\label{tab_delT}
\end{table*}

Changing the action from $\mathcal{O}(v^{4})$ to $\mathcal{O}(v^{6})$ results
in a decrease in the strength of the spin-dependent interactions. This is 
evidenced by a decrease in the mass difference between $\Upsilon$ and $\eta_{b}$
states \cite{meinel10}. For our calculation, done at a single lattice spacing of
about 0.09 fm the $\Upsilon$-$\eta_{b}$
mass difference is 56(1) MeV and 24(3) MeV for 1S and 2S states respectively using 
the $\mathcal{O}(v^{4})$ action. These values are reduced to 43(2) MeV and 14(2) MeV
when $\mathcal{O}(v^{6})$ effects are included. It is expected that inclusion of
radiative corrections to the coefficients of the NRQCD action
would raise these values somewhat \cite{hamm11}. For reference, the experimental
values are 69.3$\pm$2.8 MeV (from the Particle Data Group \cite{pdg12}) and 
59.3$\pm$1.9$^{+2.4}_{-1.4}$ MeV (from Adachi {\emph {et al}}. \cite{adachi11}) for 1S.
For 2S there are claims of 48.7$\pm$2.3$\pm$2.1 MeV \cite{dobbs12} and 24.3$^{+4}_{-4.5}$ MeV
\cite{mizuk12} for the spin splitting.

\begin{figure}
\scalebox{0.45}{\includegraphics*{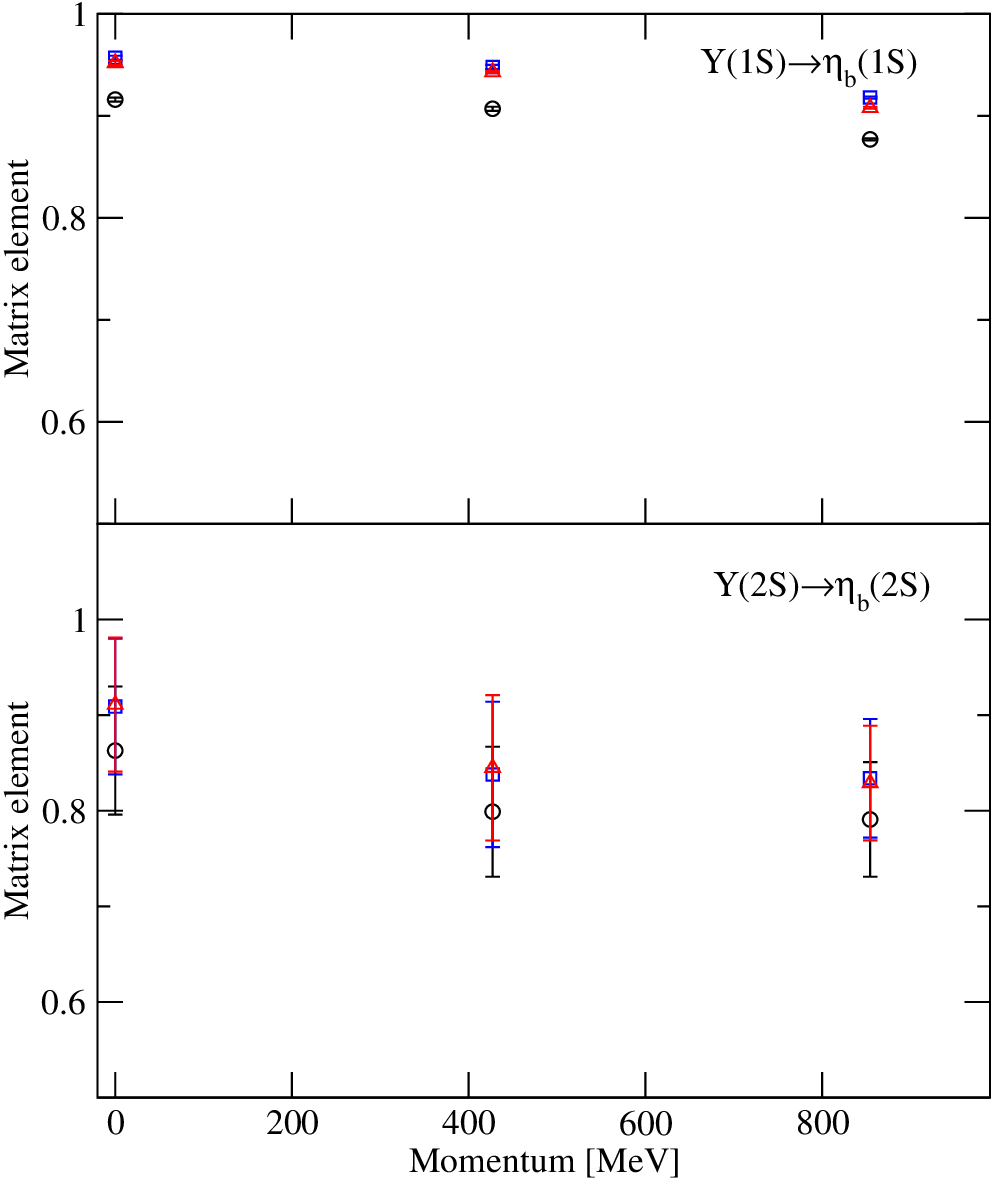}}
\caption{The matrix elements for decay of $\Upsilon$ to $\eta_{b}$ with the same
principal quantum number, as a function of momentum.
Shown are $\mathcal{O}(v^{4})$ (circles) and $\mathcal{O}(v^{6})$ (squares) action results
with the leading $\mathcal{O}(v^{4})$ magnetic operator.
Triangles include the relativistic correction to the magnetic operator in the 
$\mathcal{O}(v^{6})$ calculation.}

\label{M1momupsg}
\end{figure}

\begin{figure}
\scalebox{0.45}{\includegraphics*{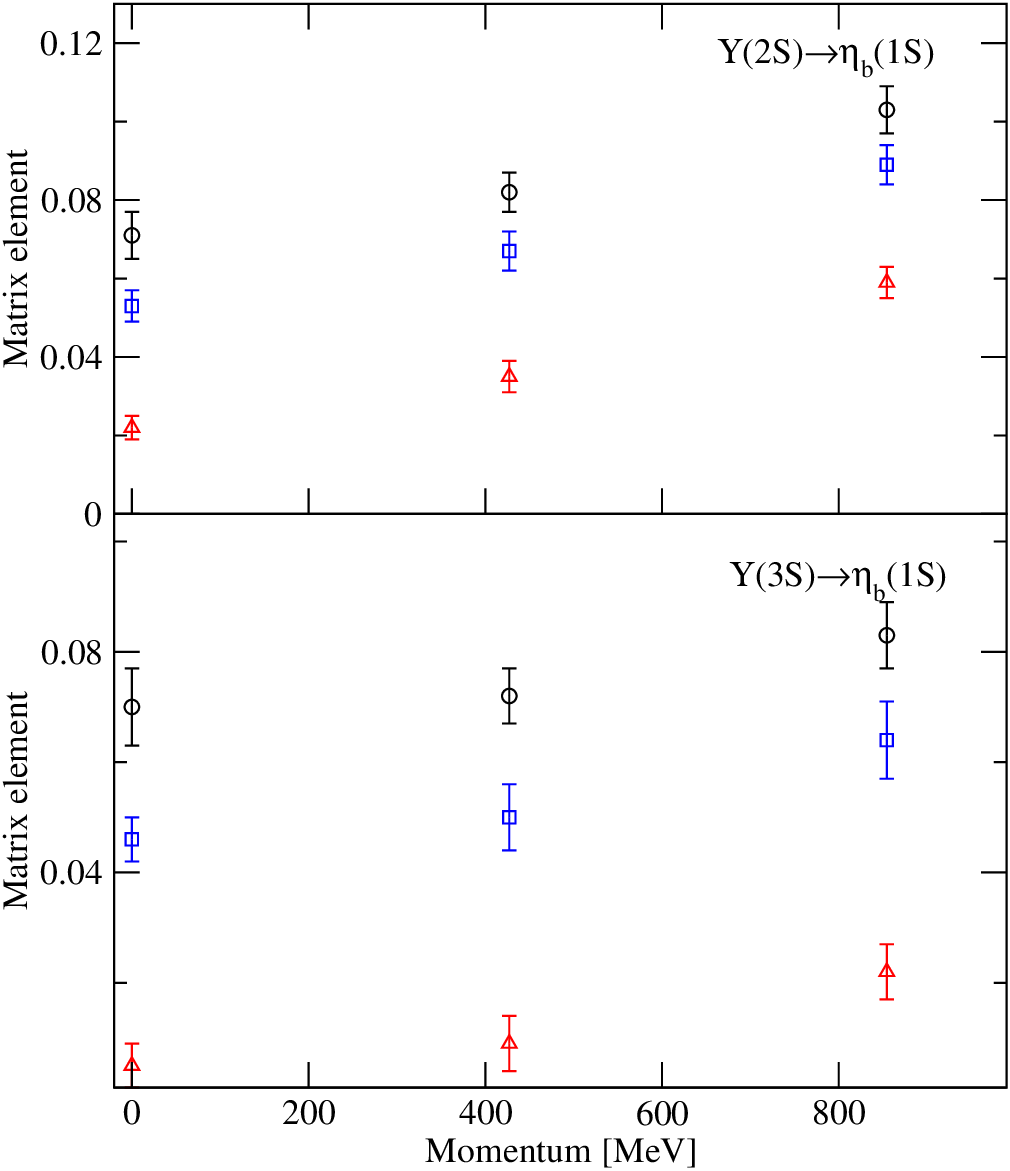}}
\caption{The matrix elements for decay of an excited $\Upsilon$ to the $\eta_{b}$ 
ground state, as a function of momentum. 
Shown are $\mathcal{O}(v^{4})$ (circles) and $\mathcal{O}(v^{6})$ (squares) action results
with the leading $\mathcal{O}(v^{4})$ magnetic operator.
Triangles include the relativistic correction to the magnetic operator in the 
$\mathcal{O}(v^{6})$ calculation.}
\label{M1momups}
\end{figure}

\begin{figure}
\scalebox{0.45}{\includegraphics*{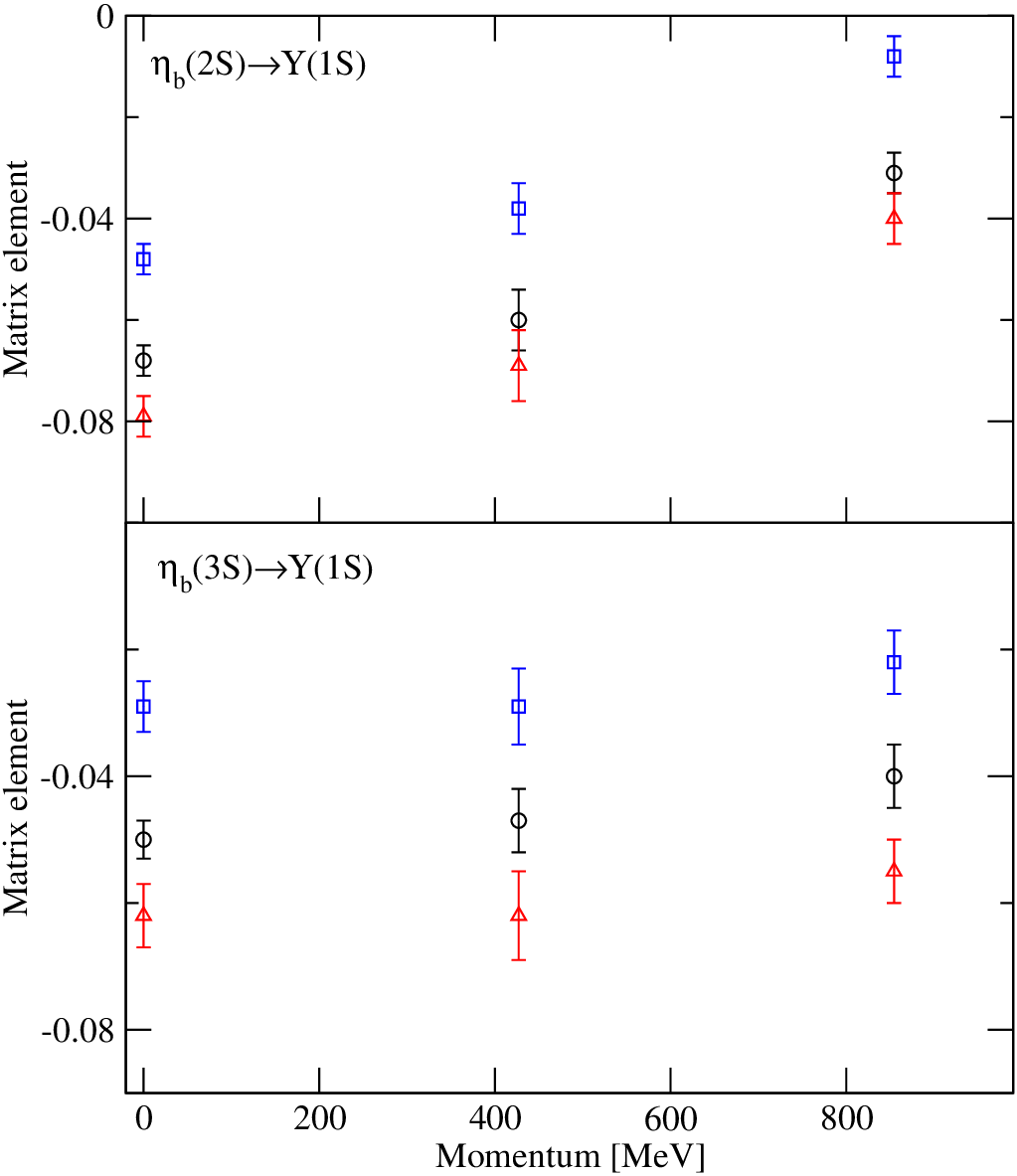}}
\caption{The matrix elements for decay of an excited $\eta_{b}$ to the $\Upsilon$  
ground state, as a function of momentum. 
Shown are $\mathcal{O}(v^{4})$ (circles) and $\mathcal{O}(v^{6})$ (squares) action results
with the leading $\mathcal{O}(v^{4})$ magnetic operator.
Triangles include the relativistic correction to the magnetic operator in the 
$\mathcal{O}(v^{6})$ calculation.}
\label{M1mometa}
\end{figure}

Decreasing the strength of the spin-dependent 
interactions leads to greater overlap of wavefunctions with the same principal quantum
number and decreases the overlap of states with different principal quantum numbers. This 
expectation is reflected in the lattice simulations. Figure \ref{M1momupsg} compares 
transition amplitudes between states with the same principal quantum number calculated 
with the $\mathcal{O}(v^{4})$ and $\mathcal{O}(v^{6})$
actions using only the leading magnetic $\mathcal{O}(v^{4})$ transition operator.
The decreased spin-dependent interaction at $\mathcal{O}(v^{6})$ shows up as an increase
of the transition amplitude. The triangles in Fig.~\ref{M1momupsg} show the effect of
adding the relativistic correction to the magnetic transition operator in the 
$\mathcal{O}(v^{6})$ calculation. The effect is very small for these transitions.

Figure \ref{M1momups} shows the comparison of the calculations for the case of excited 
$\Upsilon$ decaying to the ground state $\eta_{b}$. Here going from $\mathcal{O}(v^{4})$
(circles) to $\mathcal{O}(v^{6})$ (squares) leads to a decrease of the amplitude in
line with the notion that the overlap of wavefunctions is decreased. What is surprising is
that including the relativistic correction to the magnetic spin-flip (triangles) leads to
such a large additional decrease of the amplitude. For the decay of an excited $\eta_{b}$
to the ground state of $\Upsilon$ the situation is more complicated. The effect of spin-dependent interactions on the wavefunctions leads to a relative negative sign of the transition
amplitude compared to excited $\Upsilon$ decay. Figure \ref{M1mometa} shows that changing the 
action from $\mathcal{O}(v^{4})$ (circles) to $\mathcal{O}(v^{6})$(squares) results in a 
decrease in the magnitude of the amplitude but including the relativistic correction
to the operator (triangles) has the opposite effect.

Up to now we have discussed only the transition due to magnetic spin-flip operators.
There are additional contributions that have to be considered, that is, due to the
electric terms. These are the electromagnetic
counterparts of the spin-orbit terms in (\ref{dH4}) and (\ref{dH6}) (see also 
Eq.~(2) in Ref.~\cite{bram06}). These terms are suppressed by a factor
of (photon energy/$M_0$) relative to the magnetic  terms at the same 
order in $v^2$. They make no contribution to the transition matrix
element at zero photon momentum.  For vector to pseudoscalar transitions
between states with the same principal quantum number, where the mass difference is 
small, their effect is not important. For excited state to ground state transitions
they make a non-negligible contribution.

\begin{table*}
\caption{Three-point matrix elements for excited states decays at the physical
momentum.}
\begin{centering}
\begin{ruledtabular}
\begin{tabular}{ccccc}
&
$\Upsilon(2S)\rightarrow\eta_{b}(1S)$&
$\Upsilon(3S)\rightarrow\eta_{b}(1S)$&
$\eta_{b}(2S)\rightarrow\Upsilon(1S)$&
$\eta_{b}(3S)\rightarrow\Upsilon(1S)$\tabularnewline
&
611 MeV/$c$ &
923 MeV/$c$ &
524 MeV/$c$ &
830 MeV/$c$ \tabularnewline
\hline
$\mathcal{O}(v^{4}),\sigma$&
0.088(5)&
0.084(7)&
-0.055(4)&
-0.041(5)\tabularnewline
$\mathcal{O}(v^{6}),\sigma$&
0.073(4)&
0.067(7)&
-0.033(3)&
-0.023(5)\tabularnewline
Complete $\mathcal{O}(v^{4})$&
0.080(5)&
0.077(7)&
-0.050(4)&
-0.032(5)\tabularnewline
Complete $\mathcal{O}(v^{6})$&
0.032(5)&
0.013(7)&
-0.059(4)&
-0.041(5)\tabularnewline
Experiment&
0.034(7)&
0.018(3)&
&
\tabularnewline
\end{tabular}
\end{ruledtabular}
\par\end{centering}
\label{tab_final}
\end{table*}

Table \ref{tab_final} gives our final results for excited state decay amplitudes 
interpolated (for 2S decays) or extrapolated (for 3S decays) to the physical momentum.
The values in the first two rows were obtained using an NRQCD action with terms up
to the indicated order but only the leading magnetic spin-flip transition operator.
For the third and fourth rows all terms in the transition operator up to the order 
of those included in the NRQCD action were evaluated. The difference between the first
and third rows is due to the $\mathcal{O}(v^{4})$ electric term.
The difference between the second and fourth rows is dominated by the correction to
the magnetic spin-flip coupling. In the last row, the transition amplitude
inferred from the experimentally observed $\Upsilon(2S)$ and $\Upsilon(3S)$ 
decays \cite{babar08,babar09} is given. Our best estimate, complete $\mathcal{O}(v^{6})$,
is consistent with the experimental values. 

The excited state decays are so-called hindered transitions and the amplitudes 
depend on the interplay of spin-dependent effects on the states, relativistic
corrections and recoil (momentum) effects. The systematics of excited $\Upsilon$
and excited $\eta_{b}$ decays are different and experimental information on $\eta_{b}$ decays
would provide a very stringent test of calculations. Now that there are experimental
glimpses \cite{dobbs12,mizuk12} of $\eta_{b}(2S)$ one may hope that a measurement 
of its radiative decay may be feasible in the not too distant future. 

In this brief report we have updated a previous calculation \cite{lewis11} of bottomonium  
radiative decay amplitudes by extending the lattice NRQCD action to include spin-dependent terms
of $\mathcal{O}(v^{6})$ and by including contributions to the transition operators
that are consistent with the action.
These changes bring the calculated decay amplitudes into agreement with values determined
from observed excited $\Upsilon$ decays.  The NRQCD action and transition operators were 
constructed  using tree-level coefficients (tadpole improved for the action). Radiative 
corrections to the coefficients were not considered here. 
These have been calculated for some cases and noticeable effects were
found \cite{hart07,hamm11}. A further refinement of the calculation would have to 
apply such corrections in a complete and consistent way.

Finally, we note that the large changes in the excited state decay amplitudes found
in going from $\mathcal{O}(v^{4})$ to $\mathcal{O}(v^{6})$ NRQCD may suggest that it
would be beneficial to avoid nonrelativistic approximations altogether. The construction
and use of relativistic lattice actions for heavy quarks is the subject of ongoing
investigations \cite{yaoki12,mcneile12}. The comparison of nonrelativistic and
relativistic approaches to excited bottomonium radiative decays would be a 
worthwhile direction for future work.

\vspace{2mm}
We thank the PACS-CS Collaboration for making their dynamical gauge 
field configurations available.
This work was supported in part by the 
Natural Sciences and Engineering Research Council of Canada.


\end{document}